\documentclass[10pt]{article}
\usepackage{procl2e,epsfig}

\def\Journal#1#2#3#4{{#1}\ {\bf #2}, #3 (#4)}

\def\NPB{{\em Nucl.\ Phys.}\ B}
\def\PLB{{\em Phys.\ Lett.}\ B}

\def\PRP{\em Phys.\ Rep.}
\def\ZPC{{\em Z.\ Phys.} C}
\def\JETPL{\em JETP Lett.}
\def\SJNP{\em Sov.\ J.\ Nucl.\ Phys.}
\def\SPJETP{\em Sov.\ Phys.\ JETP}
\def\ibid{\em ibid.}
\def\citd#1#2{\cite{#1}$^{,\,}$\cite{#2}}

\def\nc{N_c}
\def\alps{\alpha_s}
\def\e{{\rm e}}
\def\apm{(1-\alpha_{\cal{P}})}

\def\be{\begin{equation}}
\def\ee{\end{equation}}
\def\bea{\begin{eqnarray}}
\def\eea{\end{eqnarray}}

\begin{document}

 \begin{flushright}
 Cavendish-HEP-96/12 \\
 hep-ph/9608250
 \end{flushright}
 \vspace{0.3cm}

\title{UNITARITY AND SATURATION IN THE DIPOLE 
 FORMULATION\footnote{Talk at DIS96, International Workshop on Deep
 Inelastic Scattering and Related Phenomena, Rome, April 1996.}
}

\author{G.P. SALAM}
\address{Cavendish Laboratory, University of Cambridge,\\
	 Madingley Road, Cambridge CB3 0HE, England}


\maketitle\abstracts{This talk reviews briefly some of the main
 results of the small-$x$ dipole formulation with regards to unitarity
 corrections. It illustrates the correspondence between unitarity and
 saturation corrections in the dipole approach and multiple
 $t$-channel pomeron exchange in the traditional BFKL view, and
 discusses how one can estimate and understand the effects of
 saturation.
}


 \noindent The dipole formulation\citd{Muel_dpl}{NZ_dpl} is an
 approach to small-$x$ physics which for total cross sections is
 equivalent to the BFKL approach\cite{BFKL} and which has been used in
 a variety of phenomenological studies.\cite{NPR95} It is reviewed in
 the talk by Webber:\cite{BRW} in the limit of a large number of
 colours, the small-$x$ heavy quarkonium (onium) light-cone wave
 function can be represented as a chain of colour dipoles stretching
 in impact parameter between the quark and antiquark. The interaction
 between onia is then due to the independent scattering of the dipoles
 in the two wave functions. To a first approximation, in any given
 collision only one dipole from each onium is involved in the
 scattering. This leads to an amplitude which violates the unitarity
 bound. The solution is to take into account multiple
 scatterings\cite{Muel_dpl} (or equivalently multiple pomeron
 exchange) and the resulting amplitude does satisfy the unitarity
 bound. The unitarity corrections are much stronger for the elastic
 cross section than for the total cross section,\cite{GPS_unt} because
 the former is dominated by more central impact parameters, where the
 amplitude, and therefore the multiple scattering corrections, are
 largest. Other significant points are (i) that the multiple-pomeron
 series does not give a convergent sum --- it is necessary to sum over
 multiple scatterings {\em before} averaging over onium configurations
 --- and (ii) that the eikonal approximation fails very badly, because
 mean properties of the onium wave function are very unrepresentative
 of the configurations which are typically involved in
 scattering.\citd{Muel_dpl}{GPS_unt}

\begin{figure}
\begin{center}
\epsfig{file=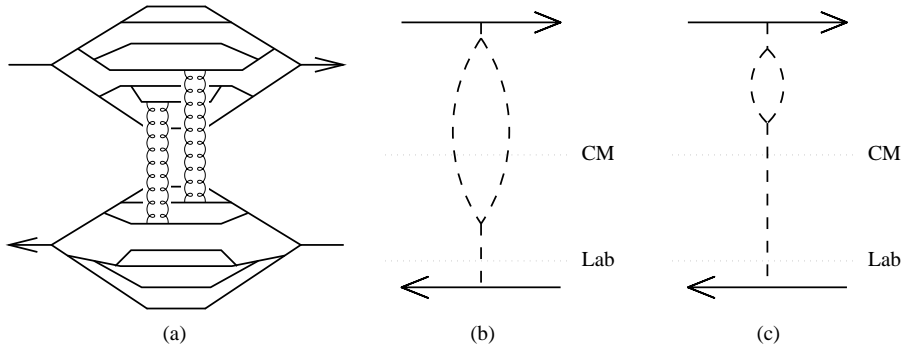, width=\textwidth}
\end{center}
\caption[]{Two-pomeron contributions to onium-onium scattering in the 
 dipole (a) and $t$-channel (b,c) views. Solid lines are dipoles,
 dashed lines are pomerons and dotted lines represent the division of
 rapidity (increasing from top to bottom) in different frames.}
\label{fig:frames}
\end{figure}

\sloppy
 The connection between the dipole and the traditional $t$-channel
 pictures of high-energy scattering is illustrated for
 centre-of-mass (CM) scattering in fig.~\ref{fig:frames}; (a) shows
 the evolution of the dipole structure (solid lines) for the squared
 wave functions of two onia. Single scattering would be caused by the
 exchange of a colour-neutral pair of gluons between a dipole in each
 onium. Here, double scattering occurs. The equivalent $t$-channel
 diagram is illustrated in (b). Everything in (b) above the dotted
 line labelled CM relates to evolution of the upper onium of (a), and
 things below it are associated with the lower onium. In the upper
 onium, the two dipoles involved in the scattering have a common
 origin early in the branching, associated with the early $1\to2$
 pomeron vertex in (b). The dipoles involved in the scattering in the
 lower onium have a common origin midway through the branching, so the
 $2\to1$ vertex in (b) occurs midway down the lower part of the
 diagram. The critical point is that going from the top down, the
 `branching out' ($1\to2$ vertex) is associated with the upper onium,
 and the `branching in' ($2\to1$ vertex) with the lower onium. Diagrams
 such as (c) where the `branching in' occurs in the upper onium are
 not included --- they would correspond to a pair of dipoles in the
 upper onium interacting with each other and are referred to as
 saturation.\cite{GLR}
\fussy

\begin{figure}
\begin{center}
\small
\tabcolsep 0.01\textwidth
\begin{tabular}{@{}cc@{}}
\epsfig{file=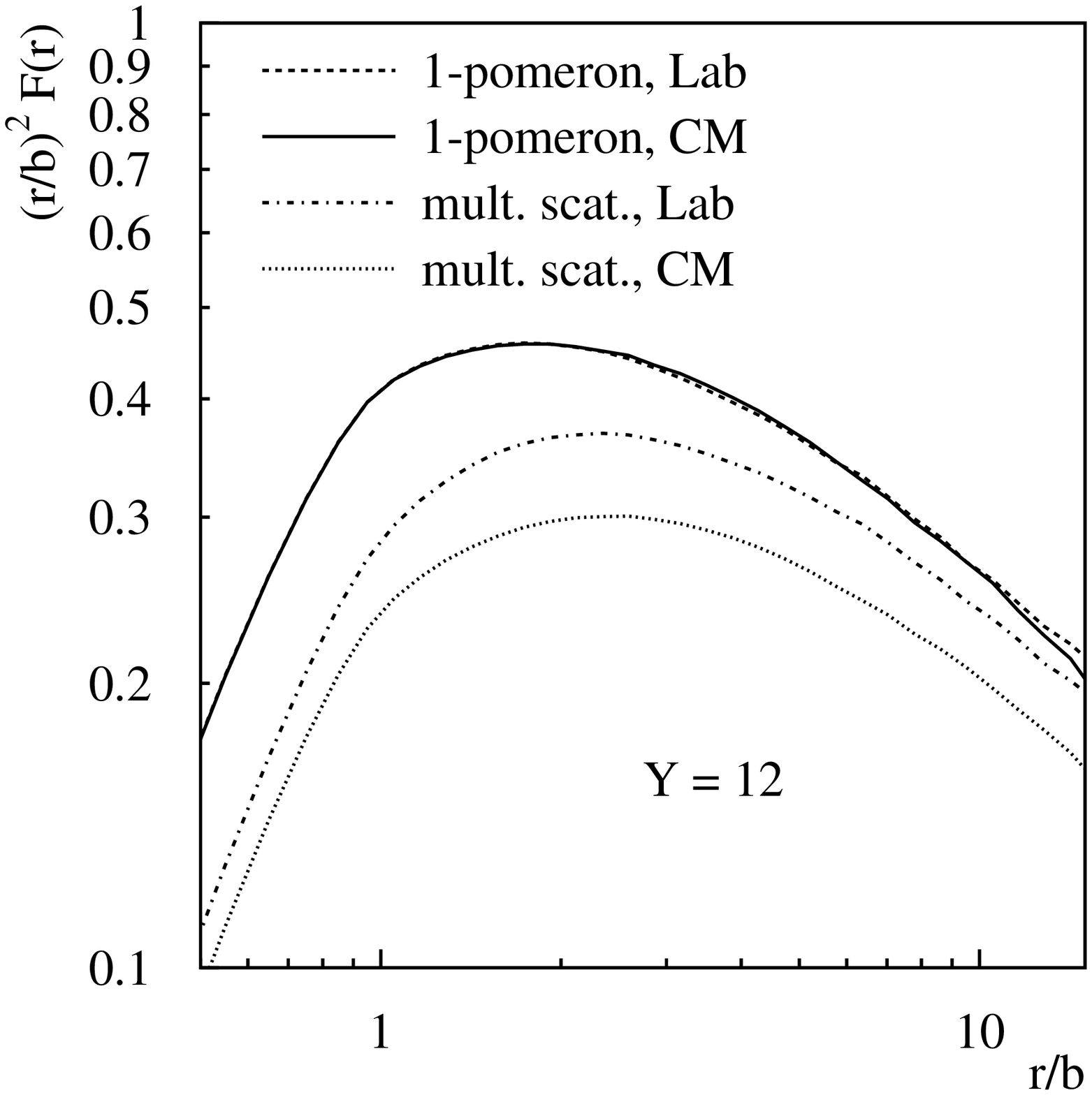, width=0.49\textwidth} &
\epsfig{file=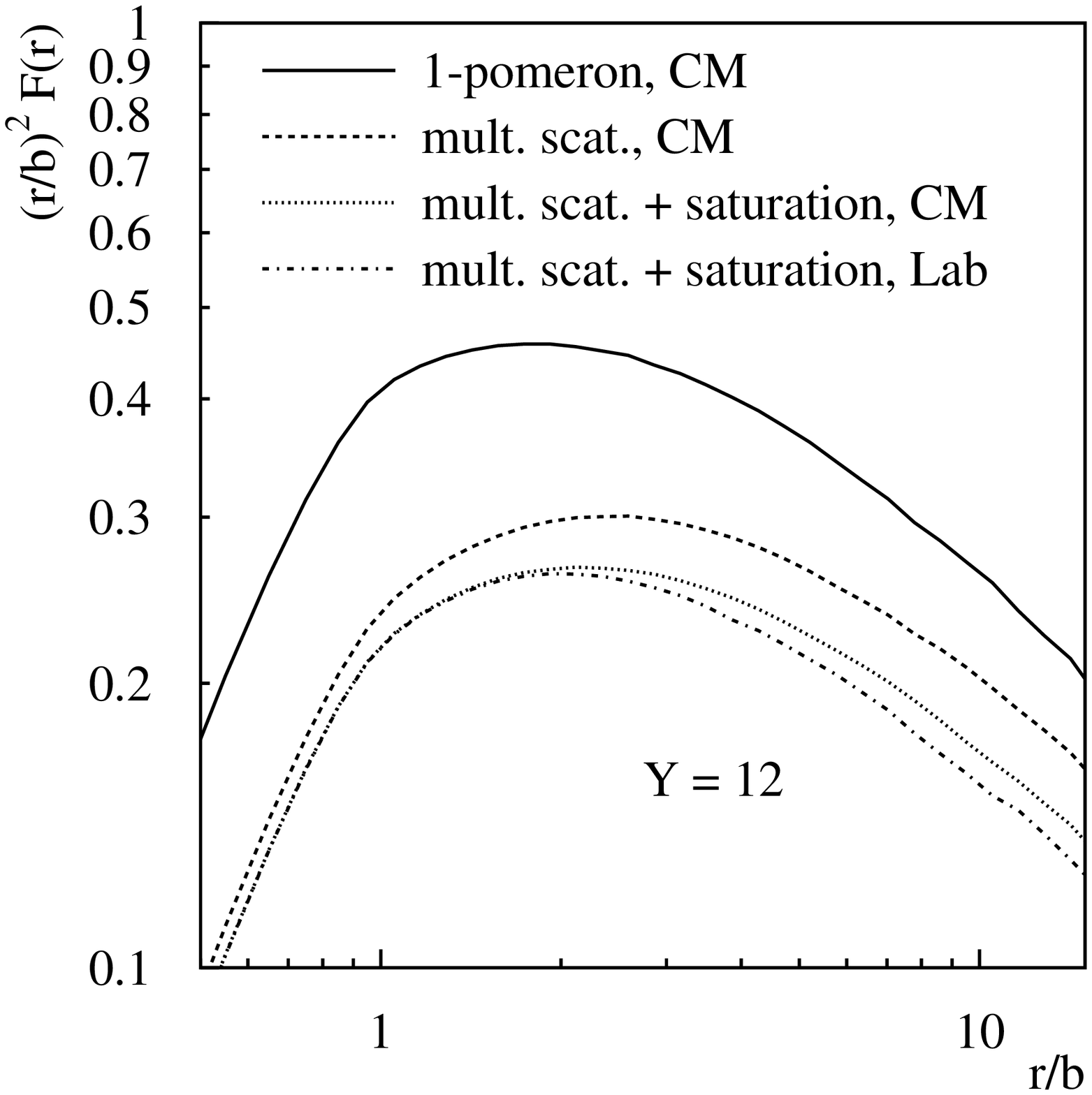, width=0.49\textwidth} \\
 (a) & (b)
\end{tabular}
\normalsize
\end{center}
\caption[]{The elastic scattering amplitude $F(r)$ for two onia of
 size $b$, separated in transverse position by a distance $r$.
 Results obtained using OEDIPUS.\citd{OED}{MS_cs}}
\label{fig:unss}
\end{figure}

 One can argue that saturation contributions can be
 neglected:\cite{Muel_dpl} the only difference between (b) and (c) is
 the rapidity range over which the two pomerons evolve. If the total
 rapidity is $Y = \ln s$, then there is a range of order $Y/2$ over
 which (c) has only one pomeron whereas (b) has two; so (c) should be
 smaller than (b) by a factor or order $\e^{-\apm Y/2}$, where
 $\apm=4\ln2 \alps\nc /\pi$ is the usual BFKL power. This is a very
 qualitative argument and it would be interesting to check its
 validity and also to obtain some understanding of the behaviour of
 the wave functions once saturation is taken into
 account. Unfortunately an equivalent of the GLR equation\cite{GLR}
 does not currently exist for the dipole approach. But a technique
 which allows one to extract considerable information about saturation
 is to make use of the requirement of Lorentz invariance of the
 scattering amplitude. In a (almost) lab frame, where the lower onium
 is (almost) stationary, the upper onium contains most of the
 evolution and graphs such as (b) of fig.~\ref{fig:frames} (saturation
 in the lab frame) are not included by a calculation of multiple
 scattering. So one expects the multiple-scattering corrections to be
 smaller in the lab frame than in the CM frame. This is borne out by
 fig.~\ref{fig:unss}a (note that for 1-pomeron exchange, the amplitude
 is frame-invariant). Roughly, the effect of saturation must just be
 to change the lab frame result so that it is equal to the CM result.

 More precisely one can make a guess for the effect of saturation on
 the dipole evolution and then tune it so as to obtain a scattering
 amplitude which is frame independent for any onium-onium scattering.
 To simulate the effect of saturation, one multiplies the evolution
 rate of a dipole $i$ by a factor $S_i(\rho_i)$, where $\rho_i$ is the
 local density of dipoles. $S_i$ is chosen so as to ensure the
 frame-independence of the scattering amplitude, and a suitable form
 is

\begin{equation}
S_i = \frac{1 - \e^{-2\Omega\alps^2\rho_i}}{2\Omega\alps^2\rho_i}
    = 1 - \Omega\alps^2\rho_i + O(\alps^4\rho_i^2),
\end{equation}

 \noindent where $\Omega$ is a constant of order $1$. Note that for
 small dipole densities, the effect of saturation is linear in the
 local dipole density (other forms which are linear work equally well
 and give similar results --- forms which are non-linear in the dipole
 density fail). Examining fig.~\ref{fig:unss}(b) one sees that with
 the inclusion of saturation, the amplitude is now the same in lab and
 CM frames (though there is a small discrepancy at large $r$).

 At small $r$, as expected, the effect of saturation is negligible in
 the CM frame. But at large $r$, the effect of saturation is as large
 as that of multiple scattering: in dipole language, what occurs is
 that multiple scattering tends to happen on the large scale where
 dipoles are dilute, whereas saturation is also affected by the
 earlier stages of the evolution, on small and moderate scales, where
 dipole densities are higher. The argument for the suppression of
 saturation given earlier did not take into account the complications
 that can arise when there is an interplay between two different
 transverse scales (in this case, $r$ and $b$).

 Nevertheless for the total cross section, which is dominated by
 moderate impact parameters, these studies suggest that saturation can
 be neglected compared to multiple scattering. But in situations where
 two different scales play a r\^ole, such as large-impact parameter
 scattering, or DIS, it is to be expected that saturation effects may
 be comparable to those of multiple scattering.

\section*{Acknowledgements}
\sloppy
 The results presented here have been obtained in collaboration with
 A.H.~Mueller.\cite{MS_cs} I am grateful to B.R.~Webber for many helpful
 discussions and to the UK PPARC for financial support.


\section*{References}
\begin{thebibliography}{99}
\bibitem{Muel_dpl} A.H.~Mueller, \Journal{\NPB}{415}{373}{1994}; 
	A.H.~Mueller and B.~Patel, \Journal{\ibid}{425}{471}{1994};
	A.H.~Mueller, \Journal{\ibid}{437}{107}{1995}; Z.~Chen and
	A.H.~Mueller, \Journal{\ibid}{451}{579}{1995}.

\bibitem{NZ_dpl} N.N.~Nikolaev and B.G.~Zakharov,
	\Journal{\ZPC}{64}{631}{1994}; N.N.~Nikolaev and
	B.G.~Zakharov, \Journal{\JETPL}{59}{6}{1994}


\bibitem{BFKL} Ya.Ya.~Balitski\v{\i} and L.N.~Lipatov,
	\Journal{\SJNP}{28}{822}{1978};  E.A.~Kuraev, L.N.~Lipatov,
	and V.S.~Fadin, \Journal{\SPJETP}{28}{822}{1978}; L.N.~Lipatov,
	\Journal{\ibid}{63}{904}{1986}.


\bibitem{NPR95} See for example N.N.~Nikolaev and B.G.~Zakharov, these
	Proceedings [hep-ph/9607479]; H.~Navelet, R.~Peschanski and
	Ch.~Royon, \Journal{\PLB}{366}{329}{1995}.


\bibitem{BRW} B.R.~Webber, 
	these Proceedings, preprint Cavendish-HEP-96/2 [hep-ph/9607441].

\bibitem{GPS_unt} G.P.~Salam, \Journal{\NPB}{449}{589}{1996};
	\Journal{\ibid}{461}{512}{1996}. 

\bibitem{GLR} L.V.~Gribov, E.M.~Levin and M.G.~Ryskin,
	\Journal{\PRP}{100}{1}{1983}.

\bibitem{OED} G.P.~Salam, preprint Cavendish-95/07 [hep-ph/9601220].

\bibitem{MS_cs} A.H.~Mueller and G.P.~Salam, preprint CU-TP-746
 [hep-ph/9605302], to appear in \NPB.

\end{thebibliography}
\end{document}